\documentclass[preprint,showpacs,showkeys,aps,prl,,groupedaddress]{revtex4-1}
\usepackage{graphicx}
\usepackage[caption = false]{subfig}
\usepackage{dcolumn}
\usepackage{bm}%

\begin{document}

\title{On the possibility of rho-meson condensation in neutron stars}

\date{\today}

\author{Ritam Mallick} 
\email[]{mallick@fias.uni-frankfurt.de}
\affiliation{Frankfurt Institute for Advanced Studies, 60438 Frankfurt am Main, Germany}
\author{Stefan Schramm}
\email[]{schramm@th.physik.uni-frankfurt.de}
\affiliation{Frankfurt Institute for Advanced Studies, 60438 Frankfurt am Main, Germany}
\author{Veronica Dexheimer}
\email[]{dexheim@kent.edu}
\affiliation{Department of Physics, Kent State University, Kent, Ohio 44242, USA}
\author{Abhijit Bhattacharyya}
\email[]{abphy@caluniv.res.in}
\affiliation{Department of Physics, University of Calcutta, 700009 Kolkata, India}

\begin{abstract}
The possibility of meson condensation in stars and in heavy-ion collisions has been discussed in the past. Here, we study whether
rho meson condensation ($\rho^{-}$) can occur in very dense matter and determine the effect of strong magnetic fields on this condensation.
We find that rho meson condensates can appear in the core of the neutron star assuming a rho mass which is reduced due to in-medium effects. 
We find that the magnetic field has a non-negligible effect in triggering condensation.
\end{abstract}



\maketitle


The study of strongly interacting matter at high densities and/or temperatures is a central topic of modern nuclear physics. 
From the experimental side the main approaches to learn about matter under these conditions are the study of relativistic heavy-ion collisions (HIC)
and observations of neutron stars, where the former approach aims at studying high-temperature and the latter one high-density and low-temperature matter.
In addition, the future experiments at GSI/ FAIR with beam energies $\leq$ 30 GeV per nucleon and the beam energy scan at RHIC aim to study the properties of matter not only at high temperature but also at
comparatively high densities. 

Increasing the density of the nuclear matter it is expected that the spontaneously broken chiral symmetry, characterized by 
a large quark condensate, is at least partially restored. However, a clear observable signature of this effect is still not 
well established. From the theory side, an early  conjecture suggested by Brown and Rho \cite{brown}, argued that
hadrons (except for the pseudo-Goldstone bosons like pions) experience a  mass reduction in nuclear matter, which is proportional to the 
in-medium quark condensate. More recent calculations, using QCD sum rules \cite{hatsuda}, quark-meson models \cite{saito} and hadronic models of vacuum 
polarization \cite{jean} also suggest such mass reduction. However, a full understanding of the restoration of chiral symmetry in dense matter remains an open and 
much-discussed topic.

More direct evidence of a potential mass reduction of vector mesons, the enhancement in the production of low invariant-mass dileptons in heavy-ion collisions has been investigated extensively, following such observations for an invariant mass around $400-500$ MeV, as measured by the CERES experiment at the CERN-SPS \cite{agakichiev}. 
Various model studies ranging from simple thermal models to more detailed transport calculations, predicted enhanced dilepton production \cite{cassing,
srivastava,koch,bratkovskaya}. One of the conjectures which explains such enhanced dilepton production in HIC, suggested the reduction of the in-medium masses of the vector 
mesons. The theoretical results were consistent with experimental data. However, later
dynamical studies hinted that the earlier studies overestimated the effect \cite{cassing1}. The enhancement in dilepton production due to vector meson mass reduction 
also differed from experiment to experiment (consistent at SPS energies but not with DLS results \cite{bratkovskaya1}). Overall the situation with respect to dilepton 
enhancement is still not unambiguously resolved.

More recent experiments of $\gamma - A$ \cite{trnka} reactions provided s clearer signature of an in-medium $\omega$ mass reduction. 
In the experiment by the
TAPS collaboration \cite{trnka} the modification of the $\omega $ in nuclei was measured in photo-production experiments, where its mass was found to be 
$m_{\omega}^* = 722 \pm 4$ MeV at $0.6\, n_0$. Similar numbers were also found in $12$ GeV photon-nuclear reaction by Naruki et al. \cite{naruki}.
The $\omega$ meson mass and quark-condensate were studied in more recent NJL model calculations \cite{huguet}, where they approximately 
recovered the B-R scaling law
using constraints from the TAPS experiment. However, for the $\rho$ meson the picture is still not clear.

At the other extreme QCD exhibits exciting physics in a strong background magnetic field, since quarks are electrically charged and can be
strongly affected by the field \cite{chernodub2}. 
It changes the 
chiral symmetry breaking by increasing the quark-condensate \cite{schramm1,shovkovy}. However, the strength of the magnetic field which 
typically can influence
QCD effects is about $eB \sim m_{\pi}^2 \sim 3\times 10^{18}$ G, where $e$ is the charge of the electron, $B$ is the magnetic field and $m_{\pi}$ denotes the 
mass of pion. Quarks and antiquarks can form condensates in the presence of strong magnetic fields. For the 
quark-antiquark bound state with light quarks, i.e. for the $\rho$ meson, strong magnetic fields can enhance the formation of a condensate.

As will be discussed further below, the key idea lies in the fact that 
a sufficiently strong magnetic field can
trigger an instability in the vacuum leading to condensates in specific quark channels, especially with respect to the charged vector mesons generating non-zero $\rho$ condensates \cite{schramm1,chernodub2}. Huge magnetic fields are generated at the periphery of heavy-ion collisions, 
where the large nuclear charges and momenta magnify the magnetic field \cite{skokov}. At LHC energies of $3.5$ TeV, the strength of the magnetic field could 
be as high as $10^{20}-10^{21}$ G \cite{schramm}.  However, such a field only exists for a very short time and is restricted to a small volume.

On the other hand, the extreme conditions discussed above can occur naturally in neutron stars (NS), 
where the nuclear density is several times the nuclear saturation density ($n_0$)
and the magnetic field can be as high as $10^{18}-10^{19}$ G in their cores \cite{ferrer,dexheimer}. The high density might reduce the $\rho$ meson mass 
due to in-medium effects and with the aid of strong magnetic field, a condensate could subsequently form. 

The possibility of pion and/or kaon condensation in neutron stars has been investigated in numerous studies in the past. 
Since it was first suggested by Migdal \cite{migdal1,migdal2}, pion condensation and its impact on NS physics became the focus of intense discussion. 
There were calculations, both supporting and challenging a condensed state in the dense cores of compact stars.
Similarly, for kaon ($K^{-}$) condensation \cite{kaplan}, it was argued that as the electron chemical potential is an increasing function of density, whereas the effective (anti-)kaon 
mass decreases with density, condensation sets in at some density. Depending on the kaon-nucleon sigma term and the specific model, kaon condensation might occur already
at about $3-4$ times saturation density, when the kaon energy becomes 
equal to the electron chemical potential, and above this density it is favourable for neutrons to decay to protons and kaons rather than
protons and electrons, giving rise to a kaon condensate. Such densities can easily be achieved in the cores of neutron stars and, therefore, 
condensation can take place. However, non-linear terms and the occurrence of hyperons tend to shift the onset of kaon condensation to higher densities \cite{Mishra:2009bp}, beyond values reached in neutron stars.

One general problem for all condensate studies is their effect on the maximum possible neutron star masses.
Since the discovery of pulsars PSR J1614-2230 and PSR J0348+0432 \cite{demorest,antonidis} with masses approximately $2 M_{\odot}$ puts severe constraints on the stiffness of the equation of state (EoS) of the core region of stellar matter. 
Therefore, a potential onset of a condensate is important since (zero-momentum) condensates soften the equation of state thereby reducing maximum star masses.

In this Letter, we study the possibility of rho meson ($\rho^{-}$) condensation occurring inside a NS and 
look into the conditions which might make rho condensation possible.
There are various aspects that increase the possibility of condensation in a neutron star environment.
Firstly, in the core of the star extreme densities exist, which could strongly amplify a potential density-dependence of the meson mass. 
Secondly, as the neutron star is in a charge-neutral state, condensation of negatively charged particles set in when their mass drops below the lepton Fermi energy, in striking contrast to the situation in a heavy-ion collision, where there is no such effect. Finally, there is a possibility of strong extended magnetic fields in at least certain classes of neutron stars like magnetars, which can generate a substantial effect for a charged spin-1 particle.

We use a standard relativistic mean field approach to study the possibility of rho-meson condensation.
Within this approach the nucleons interact through meson exchange represented by their mean field values.
The scalar meson ($\sigma$) provides the attractive force, whereas the vector mesons ($\omega,\rho$) generate repulsion. 
In order to investigate whether the rho meson will condense in a neutron star we assume a simple relation for the medium-dependent $\rho$ mass. Certainly, in more extensive calculations a fuller study of density- (or field-) dependent masses of all involved hadrons should be performed. Here,  in order to determine whether such a condensation could take place in principle, for simplicity we adopt a linear dependence  of the rho mass on the scalar field
\begin{equation}
 m_{\rho}^*=m_{\rho}-g \sigma
\end{equation}
where $g$ is a dimensionless constant and $m_{\rho}=776$ MeV is the 
vacuum mass of the rho meson. This term effectively introduces a coupling between the vector and scalar fields. 

In the presence of a background magnetic field of strength $B$, the energy level $E_{n,S_x}$
of a particle of mass $m$, charge $e$, and spin $s$ in the Landau level $n$ is given by \cite{skalozub,ambjorn,schramm,chernodub1,chernodub2}
\begin{equation}
 E_{n,S_x}  =\sqrt{p^2+m^2+(2n-g_B S_x+1)eB}
\end{equation}
where $S_x$ is the spin projection onto the magnetic field axis, and $g_B$ is gyromagnetic ratio of the particle which is taken to be $2$ for the $\rho$ meson \cite{samsonov,bhagwat,hedditch}. Therefore, the energy squared of the lowest state for a charged rho meson corresponding to $p=0,n=0$ and $S_x=1$ is 
\begin{equation}
 {m_{\rho^{-}}^2}^*=m_{\rho^{-}}^2-eB.
\end{equation}

The upper equations point to the fact that in principle, if the in-medium effect or magnetic field is high enough, the effective mass squared of the $\rho$ 
meson can be negative and condensation 
can take place. However, for neutron stars the situation is less restrictive as mentioned above, since the mass of the $\rho^{-}$ only has to fall below the electron chemical 
potential, so that electrons can be replaced by negative rho mesons for generating a charge-neutral system. 
In this case, we obtain a neutron star with rho meson condensation.

To have a better idea about if and how the rho meson condensation takes place in a neutron star, we have to adopt a specific relativistic mean field description including electrons to ensure charge neutrality of the matter. 
We choose the GM3 model for our calculation as its equation of state (EoS) has frequently been used in the literature to describe nuclear matter in NSs \cite{glendenning}. This EoS can 
generate $2$ solar mass NSs, which is an important criterion. 
We solve the mean field equations at finite density including the additional term Eq. (1). As this term does not affect isospin symmetric matter the isospin-independent quantities  of saturated matter for the
GM3 parameters do not change. However, the GM3 value of the asymmetry energy ($a_{sym} = 32.5$ MeV) changes. Therefore, while varying the field-dependence of the meson mass with the coupling $g$ from Eq. (1), we accordingly adjust the $g_{N\rho}$ coupling of the nucleon to the rho meson in order to maintain the original value of $a_{sym}$.

\begin{figure*}[ht]
\vskip 0.2in
\subfloat[]{\includegraphics[width = 4.2in]{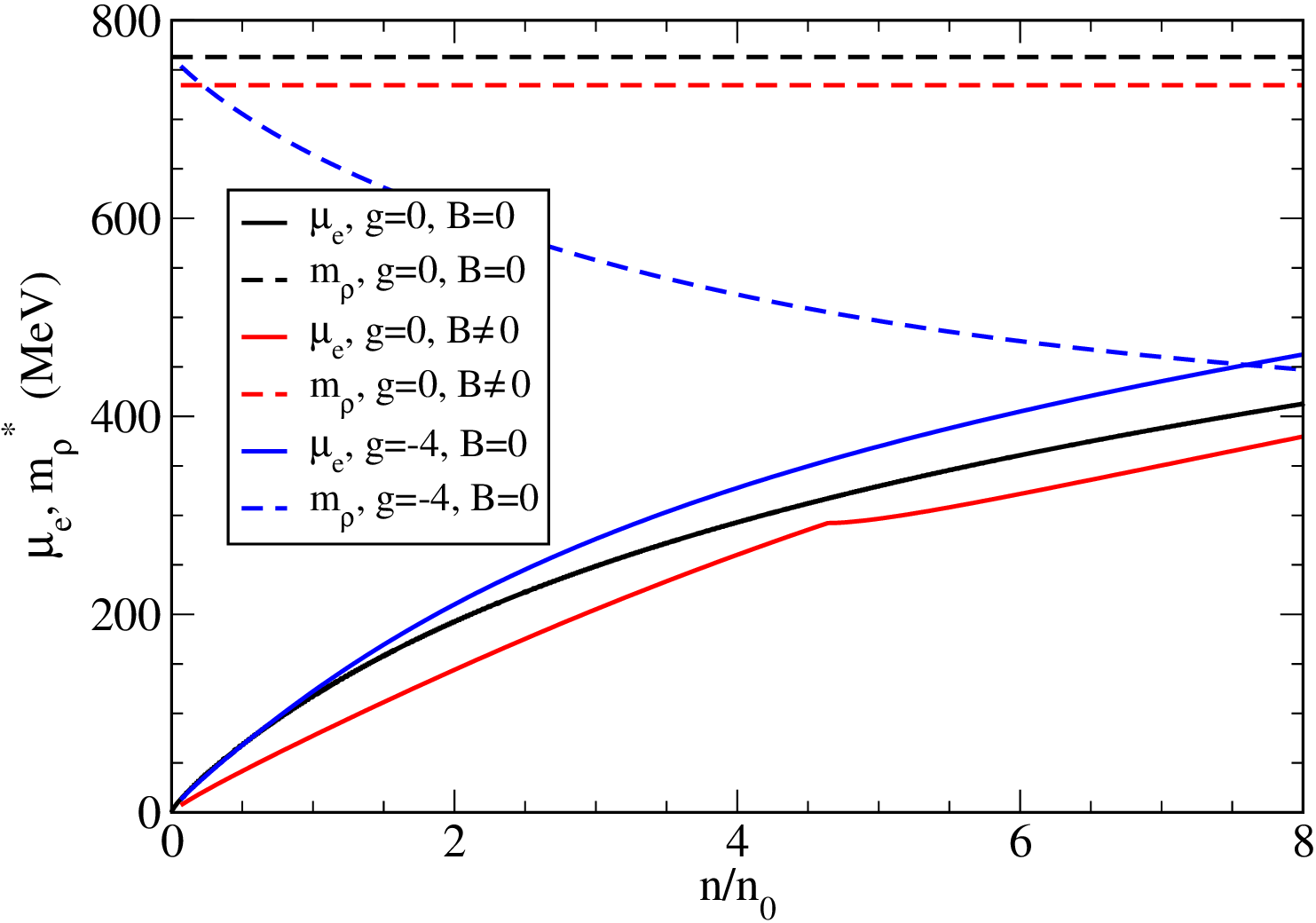}} \quad 
\subfloat[]{\includegraphics[width = 4.2in]{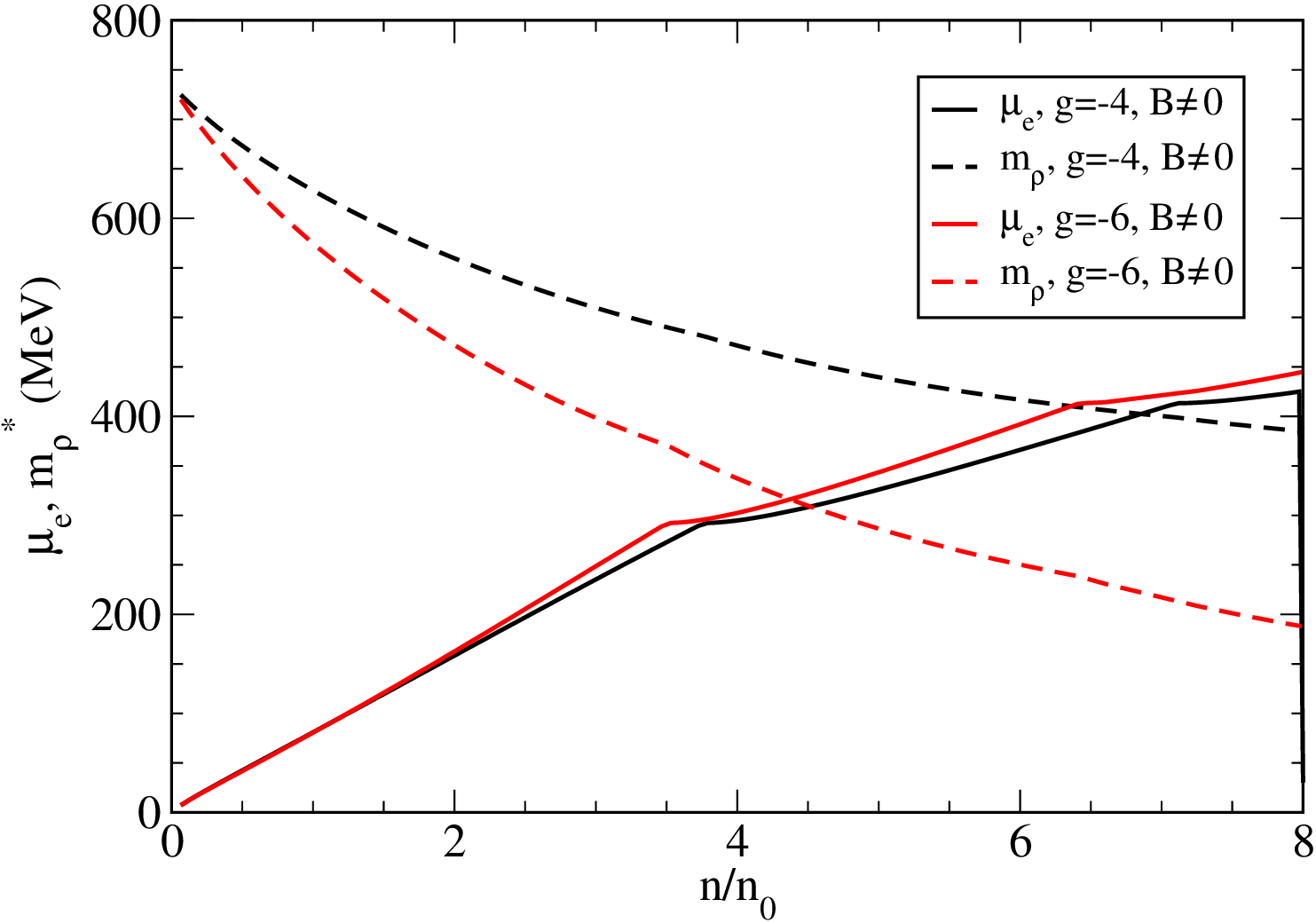}}
\caption{(Color online) The electron chemical potential (solid lines) and effective $\rho^{-}$ mass (dashed lines) are plotted as functions of normalized density. 
The crossing points mark the density at which the 
condensation appears.}
\label{fig1}
\end{figure*}

The results for stellar matter calculations are shown in Fig 1. We plot the variation of the electron chemical potential and the 
effective mass of the rho meson as a function of normalized baryon density. 
The dashed lines represent the 
rho meson effective mass and the solid lines depict the electron chemical potential. In Fig. 1a the black lines shows the respective values
without any in-medium effect or magnetic field ($g=0$, $B=0$). In this case, we find that there is no crossing point between the 
lines, which means the rho meson does not
condense in this environment for any reasonable density value. The inclusion of the magnetic field ($B\ne 0$) shifts the curves but is still 
not able to generate a crossing (red lines), which points to the fact that the
magnetic field alone cannot give rise to condensation in a compact star. In order to investigate the maximum size of the effect with some reasonable assumptions, we considered a magnetic field of a strength of $7.1\times10^{18}$ G ($5\times 10^5$ MeV$^2$). 
In order to produce rho condensation solely due to a strong magnetic field, much higher values for the magnetic field is 
required ($3-4\times10^{19}$ G). However, studies suggest that such high magnetic fields are difficult, if not impossible, to realize in a NS;  
however, they may locally appear for a short time during heavy ion collisions. For the crossing
to take place, the rho mass modification due to in-medium effect must be included. 
Taking this effect into account we find that the blue lines cross each other at about $7.6$ times nuclear saturation density. For such variation, we have assumed a vector-scalar coupling $g = -4$.
We have checked the parameter setting with nuclear matter constraints, and all are well within the accepted values. Recent
measurements of pulsar masses suggest that the central energy density of neutron stars might lie around $5-6$ times nuclear saturation values \cite{demorest}. 
In order for condensation to occur at such densities one must consider the effects of density-dependence of the mass and the effect of magnetic fields simultaneously.

In Fig 1b. we plot curves for the case in which both (in-medium and magnetic) effects are taken into account. With the above mentioned 
values of $B$ and $g$ the condensate sets in at approximately $6.9$ times saturation density. However, if we now increase the value of $g$, i.e. assuming 
a larger rho mass modification due to in-medium effects, the condensate sets in at smaller densities. For $g=-6$, the 
condensate appears around $4.4$ times nuclear saturation density. Therefore, for a neutron star with a central density of $6$ times nuclear density,
a rho meson condensate appears in the core of the star and extends up to the radius where the nuclear density is $4.4$ times nuclear 
density.

\begin{figure*}[ht]
\vskip 0.2in
\includegraphics[width = 4.2in]{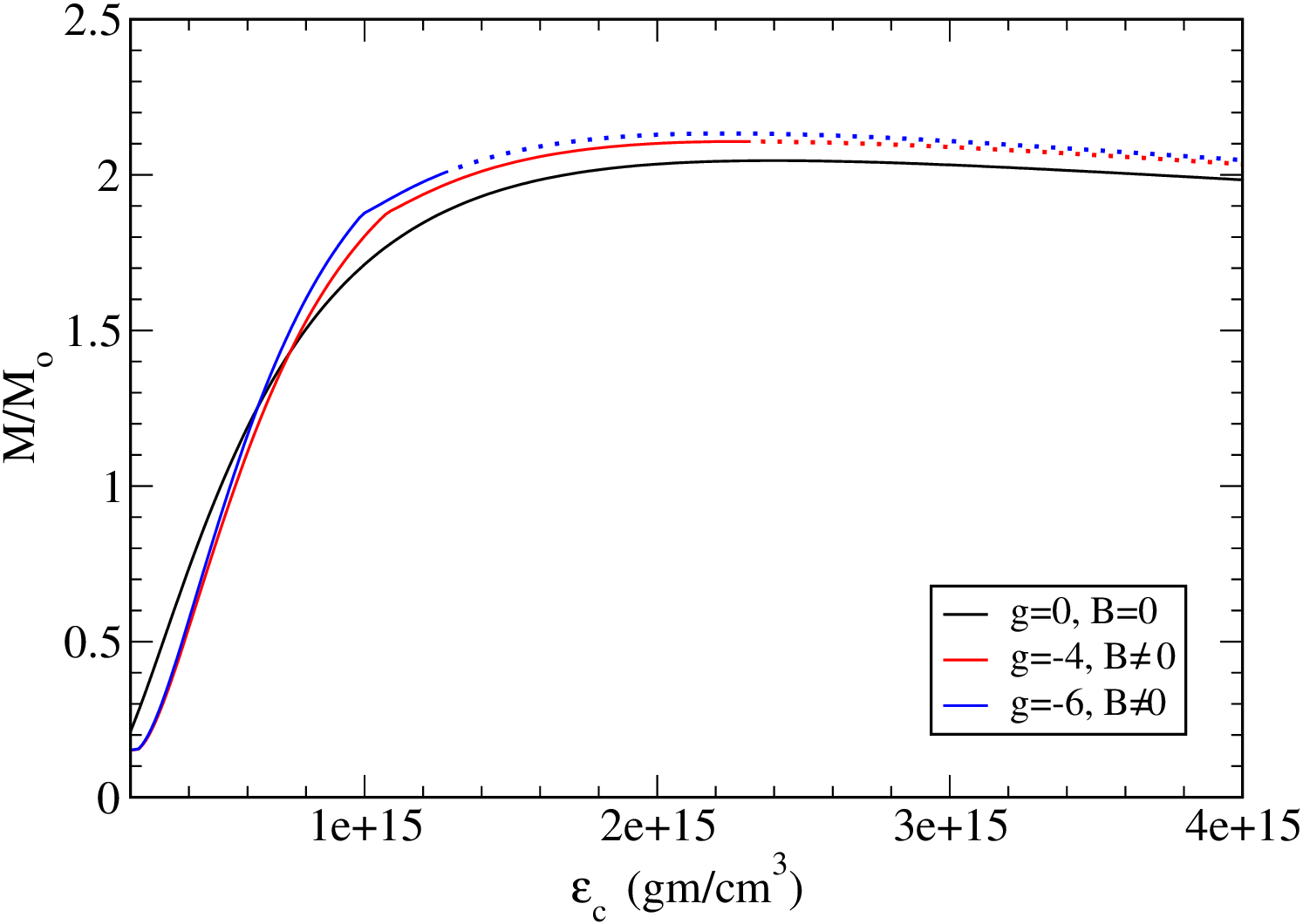}
\caption{(Color online) Sequence of stars for the GM3 EoS. The mass is shown as function of the central energy density. When both $g$ and $B$ are set to zero there is no rho condensate inside the stars 
and the maximum mass achieved is 
$2.05$ solar mass. When $g, B \ne 0$ the condensate appears. The dotted lines indicate stars that contain some amount of rho condensate.}
\label{fig2}
\end{figure*}
 
To study how rho condensation impacts the properties of a neutron star, we solve the Tolman-Oppenheimer-Volkov equation.
In Fig. 2, we plot the sequence of stars obtained using the GM3 EoS. The black line represents the results for a NS 
where there is no $\rho^{-}$ condensate. The 
maximum mass achieved is about $2.05 M_{\odot}$. The red and blue curves show the sequences where rho meson condensation takes place.
The solid lines in the curve represent stars which do not contain rho condensate, whereas the dotted points of the curve indicate stars that have 
some amount of $\rho^{-}$ condensate in their cores. For these curves the magnetic field is kept constant at $B=7.1\times10^{15}$ G. 
We find that as we increase $g$, the condensate appears much earlier i.e. more and more stars of the sequence have 
a condensate region in their cores. At this point we have determined the onset of condensation but so far have not performed a full calculation of the star structure including a condensate. However, assuming a drastic softening of the 
equation of state, the biggest possible shift of the maximum mass is given by the star mass when condensation starts to set in, which, for this specific model, yields a shift downward by about $0.2 M_{\odot}$ for a coupling of $g=-6$ and including magnetic fields as can be inferred from Fig. 2.

To summarize, we studied the possibility of rho meson condensation within a standard relativistic mean field approach.
While not rigorously proving the existence of the condensate, we have shown that such a state can be achieved, especially in a stellar environment,
using quite reasonable assumptions.
We conclude that the appearance of the condensate requires a modification of rho meson mass in the dense medium.
In addition, magnetic fields have a non-negligible effect in reducing the threshold for condensation further.
For that the magnetic field has to be quite large, at least in the high-density core region of the star; 
however, the exact upper limit of potential magnetic fields in the interior of a neutron star is still an open discussion. 

Note that although we had to use a specific model for obtaining numerical results all the arguments are quite generic and
hold as well for other models.
The actual numbers (condensation critical density, maximum mass shift due to condensation) will vary somewhat, but 
the qualitative nature of our result remains the same. One interesting consequence of the condensate would be the 
modification of the neutron star cooling, which has to be studied in more detail in the future.
On the modelling side of this work more general and microscopic, quark-level based calculations of possible vector 
meson mass changes with density and their effects on the equation of state will be presented in subsequent publications.

\end{document}